\documentclass[a4paper,11pt]{article}
\usepackage[utf8]{inputenc}
\usepackage{authblk}
\usepackage{authorindex}
\usepackage{microtype}
\usepackage{amsmath}
\usepackage{amsfonts}
\usepackage{amssymb}
\usepackage{mathrsfs}
\usepackage{bbm}
\usepackage[english]{babel} 
\usepackage[compress]{cite}
\usepackage{eufrak}
\usepackage{bmpsize-base}
\usepackage{fullpage}
\usepackage{setspace}
\usepackage{booktabs}
\usepackage{eso-pic}
\usepackage{shapepar}
\usepackage{enumitem}
\usepackage{braket}
\usepackage{graphicx}
\usepackage{epstopdf}
\usepackage[margin=10pt,font=small,labelfont=bf]{caption}
\usepackage{float}
\usepackage{subfigure}
\usepackage{bm}
\usepackage{makeidx}
\usepackage{placeins}
\usepackage{listings} 
\usepackage{color} 
\usepackage{titlepic}
\usepackage{hyperref}
\usepackage{bbm}
\usepackage{mathrsfs}
\usepackage{lineno}

\renewcommand{\epsilon}{\varepsilon}

\newcommand{\U}{\mathcal{U}}

\renewcommand{\hat}{\widehat}

\renewcommand{\'}{\`}
\renewcommand{\S}{\mathbf{S}}

{\left\lbrace\begin{array}{@{}l@{}}}%
{\end{array}\right.}

\title{The replica symmetric solution for Orthogonally Constrained Heisenberg Model on Bethe lattice}
\date{}
\author{Francesco Concetti}
\affil{\textit{Dipartimento di Fisica, Sapienza Università di Roma, Piazzale Aldo Moro 2, I-00185 Rome, Italy}}
\begin{document}

\maketitle

\begin{abstract}
In this paper, we study the thermodynamic properties of a system of $D$-components classical Heisenberg spins lying on the vertices of a random regular graph, with an unconventional first neighbor non-random interaction $J(\S_i\cdot\S_k)^2$. We can consider this model as a continuum version of anti-ferromagnetic $D$-states Potts model. We compute the paramagnetic free energy, using a new approach, presented in this paper for the first time, based on the replica method. Through the linear stability analysis, we obtain an instability line on the temperature-connectivity plane that provides a bound to the appearance of a phase transition. We also argue about the character of the instability observed.
\end{abstract}

\section{Introduction}
The study of spin glasses on locally tree-like lattices, with finite coordination number, has attracted a large interest in statistical physics since the eighties. 

At the beginning of the past decade, Mézard and Parisi \cite{ParMez,23col} adapted the replica symmetry breaking (RSB)\cite{VPM,FH} scheme to solve models on such sparse lattices, improving the Bethe-Peierls method in the cavity method. The cavity method enabled a large theoretical activity about such class of models in the past decade and important progress has been achieved. This theoretical activity was mainly concentrated on the zero-temperature limit \cite{23col} of various
constrained models on random graphs, because of the connection between such problems of statistical mechanics with this hard combinatorial optimization
problems \cite{MM} such as the K-satisfiability problem \cite{22col,24col} or the graphs coloring \cite{Zecchina,Zecchina2,Kraz0,Kraz}.

In contrast to the rich literature about discrete spins (Potts or Ising) models, few results have been achieved with vector spin glass models. 

In this paper we study a model of classical Heisenberg spins with an unconventional non-random interaction. The Hamiltonian is given by
\begin{equation}
\label{ham}
H_{G}[\S]=\sum_{\braket{ik}_G}J \big(\S_i\cdot \S_k\big)^2\, ,
\end{equation}
where $J$ is a positive coupling constant and the sum is over all edges of a given graph $G$. 

The spins $\S_i$ are $D$-dimensional vectors, lying on the unit sphere:
\begin{equation}
\S_i=(S_{i,1},S_{i,2},\cdots, S_{i,D})\,,\quad \|\S_i\|=1\,. 
\end{equation}

At zero temperature, two interacting spins tend to orientate orthogonally to each other, so we call this model Orthogonally Constrained Heisenberg Model (OCHM). 
A satisfiability version of the problem, called Orthogonal vector coloring, was studied in mathematics by Haynes, Gerald, et al. \cite{math} for generic graphs. 

The case when the graph $G$ is a finite dimensional periodic lattice was extensively investigated via numerical simulations \cite{Victor1,Victor2}. For generic non-periodic graphs $G$, the model may show a completely different physical behavior. 

In this paper, we consider the model over a simple graph belonging to the random regular graph ensemble $\mathcal{G}_{N,K}$: this ensemble consists of the set of all graphs the graph with N-vertices, such that the number of links connected to each vertex is equal to $K$ \cite{RG}. In spin glass literature, such ensemble of random graphs is considered as a possible definition of Bethe lattices \cite{ParMez}.

Because edges are random, for large $N$ and $K>2$, the typical size of a loop is of order $\log N$, so the probability to have finite loops vanishes in the $N\to \infty$ limit. As consequence, random regular graphs (RRG) are locally isomorphic to a tree: such models are exactly solvable in mean field theory. 

Because the coordination number is constant for each vertex, at finite length scale RRG do not show any disorder. The random nature of RRG enters at the global level because of the presence of large loops. Large loops can induce frustration, and the model actually behaves as a mean-field spin glass.

\begin{figure}[h!]
\centering
{\includegraphics[width=0.5\textwidth]{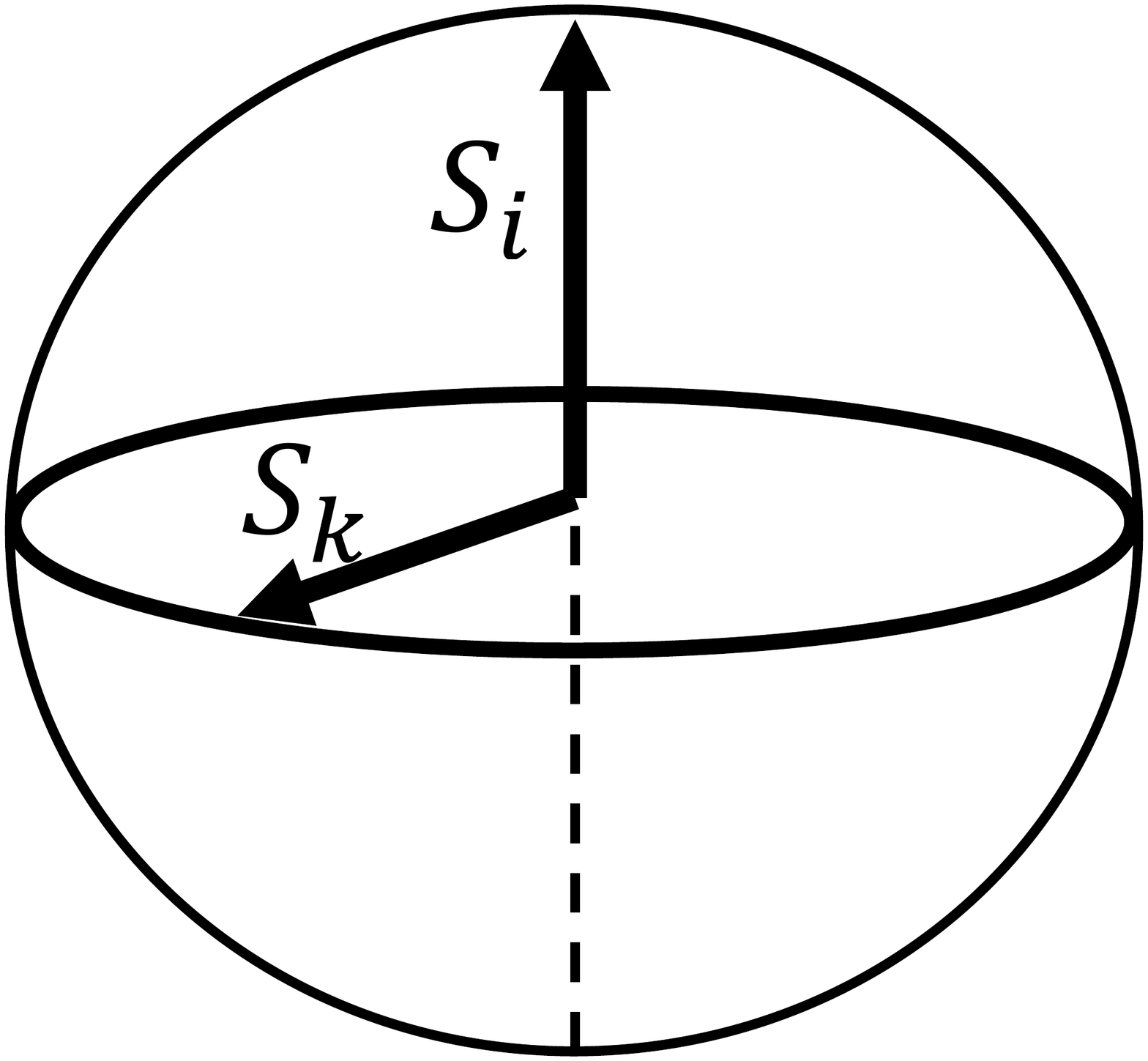}} \quad
\caption{\small The spin $S_i$ is quenched at the pole and $S_k$ can range over all the equator.}
\label{sphere}
\end{figure}

For $D\geq 3$, unlike the Ising spin glass model defined on the same ensemble of graphs\cite{ParMez,M,G2}, the condition of minimum energy for each couple of first neighbor spins is infinitely degenerate (Fig.\ref{sphere}). This feature, together with the non-randomness of coupling constants, makes this model quite similar to well-known anti-ferromagnetic $D$-states Potts model (AF $D$-PM) defined on RRG, where a degeneracy in single pair interaction also occurs \cite{Wu}. We can consider AF $D$-PM as a discrete version of $D$-components OCHM where spins are quenched in $D$ orthogonal fixed possible directions: each direction corresponds to a Potts state.

From this analogy, we can argue that, if the AF $D$-PM on a given graph is not frustrated, i.e. the graph is $D$-colorable \cite{Zecchina,Zecchina2,Kraz0,Kraz}, the OCHM on the same graph is not frustrated too \cite{math}. 

Whilst, in Ising spin glasses, a competition between random ferromagnetic and anti-ferromagnetic like interactions is necessary for the formation of a spin glass phase \cite{VPM,FH}, in the presented model, as in AF $D$-PM defined in the same graphs, the interaction is non-random and every edge has locally the same environment. For this reason, OCHM and AF $D$-PM constitute a different class of spin glasses, where the possibility to get a spin glass phase arises from the degeneracy of two spins minimum interaction energy level, that allow the system to form a self-generated disorder. 

The paper is organized as follows: in section \ref{2} we obtain a single vertex free energy functional using the replica formalism and then in section \ref{3} we obtain, from variational principle, the mean field self-consistency equation for a proper order parameter; in section \ref{3} we also compute the Stability operator (Hessian) for a general solution of the mean field equation; in section \ref{4} we compute analytically the free energy for the rotational invariant (paramagnetic) phase and then we perform the study of the stability of such solution, and analyze the results. 
\section{The replicated free energy functional}
\label{2}
In this section we formulate the problem in a variational way: we define a variational effective free energy functional depending on a suitable order parameter. 

For each given graph $G$, we may define the partition function and the free energy per spin:
\begin{gather}
\mathcal{Z}_G=\int \prod_i^N(d\Omega_{S_i})\exp(-\beta H_{G}[\S] )\, ,\\
f_G=\frac{1}{N\beta}\log Z_G\, ,
\end{gather}
where $d\Omega_{S_i}$ is the angular differential element for the spin $\S_i$.

We shall perform the quenched average over all the allowed graphs $G\in\mathcal{G}_{N,K}$ using the replica method \cite{EA1}
\begin{equation}
\label{quenchedrep}
\mathcal{Z}_n=\overline{(\mathcal{Z}_G)^n}=\frac{1}{|\mathcal{G}_{N,K}|}\sum_{G\in \mathcal{G}_{N,K}} (\mathcal{Z}_G)^n\, ,
\end{equation}
\begin{equation}
\label{f}
f=\overline {f_G}=-\lim_{N\to\infty}\lim_{n\to 0} \frac{1}{N\beta}\,\partial_n \mathcal{Z}_n\, .
\end{equation}
 The symbol $\overline{\,\cdot \,}$ stands for the average over the ensamble $\mathcal{G}_{N,K}$ and $|\mathcal{G}_{N,K}|$ is the $\mathcal{G}_{N,K}$ set cardinality, that in large $N$ limit, is given by \cite{Bender,Bollobas}
\begin{equation}
\label{labelled}
|\mathcal{G}_{N,K}|\sim e^{-(K^2-1)/4}\frac{(NK-1)!!}{(K!)^N}=\frac{C_{N,K}}{(K!)^N}\bigg(\frac{NK}{e}\bigg)^{\frac{NK}{2}},
\end{equation}
where $C_{N,K}$ is a correction factor such that $(C_{N,K})^{1/N}\to 1$ when $N\to \infty$, than it constitutes an irrelevant contribution to quenched free energy given by \eqref{f}.

As usual in replica framework, the average over disorder (the graphs) is taken into account through an effective interaction amongst different replicas variables, leading to an effective non-random mean field free energy functional in replica space \cite{VPM,FH}.

In order to compute the average \eqref{quenchedrep} we propose a nice interpretation of the problem in term of a diagrammatic theory. 

First of all, we rewrite the averaged replicated partition function \eqref{quenchedrep} in term of single vertex functions. This is the starting point for the approach we shall present in the next subsection.

For a given graph $G$, the replicated partition function $(Z_G)^n$ has the form of a product of two vertices functions:
\begin{gather}
(\mathcal{Z}_G)^n=\prod_{\braket{i,j}_G} \Psi \big(\S^{(n)}_i,\S^{(n)}_j\,\big) \, ,\\
 \Psi \big(\S^{(n)}_i,\S^{(n)}_j\,\big)=\exp\Big(-\beta\sum_{a=1}^n (\S^a_i\cdot \S^a_j)^2\, \Big)\, ,
\end{gather}
where the product is over the edges of the graph $G$ and $\beta=J/k_bT$ .

Using the Hubbard-Stratonovich identity \cite{Hubbard}, we can decouple two vertices functions in a convolution of two single vertex functions introducing, for each edge $\Braket{i,j}_G$ of the graph $G$, $n$ Gaussian averaged $D\times D$ matrices $\Xi^a$
\begin{multline}
\label{uno}
 \exp\Big(-\beta \sum_{a=1}^n(\S_i^a \cdot \S_j^a)^2\Big)=\\
\bigg(\frac{ \beta }{2\pi}\bigg)^\frac{nD^2}{2}e^{-n\beta}\int \bigg(\prod_{a=1}^n d[\Xi ^a]e^{-\frac{\beta}{2} Tr[\,(\Xi^a)^2]}\,\bigg) \exp\Big(\, i\beta\sum_{a=1}^n\sum_{\mu =1}^D\sum_{\nu=1}^D\Xi_{\mu\nu}^a(S_{i,\mu}^aS_{i,\nu}^a + S_{j,\mu}^aS_{j,\nu}^a)\,\Big)\\
=\int d^n[\Xi]d^n[\Xi ']\delta[\Xi^{(n)}-\Xi '^{(n)}]\Phi(\Xi^{(n)},\S_i^{(n)})\Phi(\Xi '^{(n)},\S_j^{(n)})\, ,
\end{multline}
where the single vertex function is:
\begin{equation}
\label{phi}
\Phi(\Xi^{(n)},\S_i^{(n)})=\bigg(\frac{ \beta }{2\pi}\bigg)^\frac{nD^2}{4}e^{-\frac{n\beta}{2}} \exp\bigg(-\frac{\beta}{4} \sum_{a=1}^nTr\big((\Xi^a)^2\big) + i\beta\sum_{a=1}^n\sum_{\mu \nu}\Xi_{\mu\nu}^aS_{i,\mu}^aS_{i,\nu}^a\bigg)\, .
\end{equation}
The symbols $\Xi^{(n)}$ and $\S^{(n)}$ stands respectively for the $n$ matrices list $\big(\Xi^1,\Xi^2,\cdots, \Xi^n\big)$ and the $n$ spins list $\big(\S^1,\S^2,\cdots, \S^n\big)$ and 
\[
d^n[\Xi] d^n[\Xi ']\delta[\Xi^{(n)}-\Xi '^{(n)}]=\prod_{a=1}^n\prod_{\mu=1}^D\prod_{\nu=1}^Dd\Xi ^ad\Xi '^a_{\mu,\nu}\delta(\Xi^a_{\mu,\nu}-\Xi'^a_{\mu,\nu})\, .
\]
We shall also use the following symbol for the differential angular elements of $n$ spin replicas:
\[
d^n\Omega_{S}=\prod_{a=1}^n d\Omega_{S^a}\,.
\]
By this formalism we have:
\begin{equation}
\label{Zrep}
Z_n=\int \Big(\prod_{i=1}^N d^n\Omega_{S_i}\Big) P[\S_1^{(n)},\S_2^{(n)},\cdots,\S_N^{(n)}]\, 
\end{equation}
with
\begin{multline}
\label{P}
P[\S_1^{(n)},\S_2^{(n)},\cdots,\S_N^{(n)}]=\\
\frac{(K!)^N}{C_{N,K}}\sum_{G=\mathcal{G}_{N,K}} \prod_{\braket{ij}_G} \int d^n[\Xi]d^n[\Xi ']\frac{e}{N}\delta[\Xi^{(n)}-\Xi'^{(n)}]\Phi(\Xi^{(n)},\S_i^{(n)})\Phi(\Xi '^{(n)},\S_j^{(n)})\,.
\end{multline}
\subsection{Diagrammatic representation and free energy functional}
In this subsection we shall propose an interpretation of the average over all RRG in term of a diagrammatic theory. In this way we shall obtain a variational single vertex free energy functional in the replica space.
 
We start recalling some fundamental properties of Gaussian expectation values, that form the basis of our approach.

Let us consider the following functional integral:
\begin{equation}
\label{pathint}
\mathcal{I}=\int \mathcal{D}[\eta] \exp\big\{ -A[\eta]\,\big\}\prod_{i=1}^N\bigg(\int d^n[\Xi]\eta(\Xi^{(n)}\,)\psi_i(\Xi^{(n)})\,\bigg)^{K}.
\end{equation}
where $A[\eta]$ is a quadratic positive functional, i.e. $\mathcal{D}[\eta] \exp\big\{ -A[\eta]\big\}$ is a Gaussian functional measure, defined on a proper space of real functions insisting on $\mathbb{R}^{n\times D\times D}$, such that:
\begin{equation}
\int \mathcal{D}[\eta] \exp\big\{ -A[\eta]\,\big\}=1\, .
\end{equation}
The list $\{\psi_i\}_{1\leq i \leq N}$ is a given set of $N$ functions depending, in some way, on the indices $i$. Supposing $NK$ is even, from Wick's theorem, the Gaussian expectation value \eqref{pathint} can be expressed as the sum over all possible ways to contract in pairs the $NK$ functions in the product $\prod_{i=1}^N\bigg(\int d^n[\Xi]\eta(\Xi^{(n)}\,)\psi_i(\Xi^{(n)})\,\bigg)^{K}$. Because the dependence on the indices $i$ of the $\psi$-s, the $NK$ functions merge in $N$ groups of $K$ functions, labeled by $i$. 

A pair contraction involving two functions, depending on the indices $i$ and $j$, is a double indices factor $\Psi_{ij}$ defined as
\begin{multline}
\Psi_{i,j}=\int d^n[\Xi]d^n[\Xi ']\bigg[\int \mathcal{D}[\eta] \exp\big\{ -A[\eta]\,\big\}\eta(\Xi^{(n)}\,)\eta(\Xi '^{(n)}\,)\bigg]\psi_i(\Xi^{(n)})\psi_j(\Xi '^{(n)})\\
=\int d^n[\Xi]d^n[\Xi ']\Delta_A(\Xi^{(n)},\Xi '^{(n)})\psi_i(\Xi^{(n)})\psi_j(\Xi '^{(n)})\, ,
\end{multline}
where $\Delta_A(\Xi^{(n)},\Xi '^{(n)})$ is the propagator associate to the quadratic functional $A[\eta]$. 
 
For a particular contracting procedure we can associate a diagram where each vertex represents one of $N$ different functions $\psi_i$ and edges are the contractions $\Psi_{i,j}$: in this way we perform the configuration model of random $K$-regular multigraphs (RRM) with $N$ vertices \cite{Bollobas}.We denote this ensemble of multigraphs with $\mathcal{G}^{\star}_{N,K}$. The summation over all possible contracting procedures is rewritable as a summation over its elements. 

In $\mathcal{G}^{\star}_{N,K}$ both simple RRG and graphs containing self-loops and multiple edges (non-simple graphs) are considered.

Some distinct contracting procedures have the same diagrammatic representation, so multigraphs arise with different frequencies, then, for each of them, we have to consider the multiplicity factor $M_{G^{\star}}$, i.e. the number of different ways the multigraph $G^{\star}$ can be assembled:
\begin{equation}
\label{multiplicity}
\mathcal{I}=\sum_{G^{\star}\in\mathcal{G}^{\star}_{N,k}} M_{G^{\star}}\prod_{\Braket{i,j}_G}\Psi_{i,j}=(K!)^N\sum_{G^{\star}\in\mathcal{G}^{\star}_{N,k}} m_{G^{\star}}\prod_{\Braket{i,j}_G}\Psi_{i,j}.
\end{equation}
In \eqref{multiplicity} we have introduced a reduced multiplicity factor $m_{G^{\star}}=M_{G^{\star}} (K!)^{-N}$. For large $N$, the multiplicity factor for a typical simple graph ${G^{\star}}$ is $m_{G^{\star}}=1$, i.e. simple graphs are uniformly distributed in $\mathcal{G}^{\star}_{N,k}$ .

If we put $\psi_i(\Xi^{(n)})=\Phi(\Xi^{(n)},\S^{(n)}_i)$, defined on \eqref{phi}, and ${\Delta_A(\Xi^{(n)},\Xi'^{(n)})=\dfrac{e}{NK}\delta[\Xi^{(n)}-\Xi'^{(n)}]}$, that is $A[\eta]=\frac{NK}{2e}\int d[\Xi^{(n)}] \eta^2(\Xi^{(n)})$, we obtain a function $P^{\star}[\S_1^{(n)},\S_2^{(n)},\cdots,\S_N^{(n)}]$ quite similar to the function defined in \eqref{P}, apart from the fact that the sum over graphs considers all RRM, biased according to configuration model. 

We can define the replicated partition function averaged on the ensamble $\mathcal{G}^{\star}_{N,K}$
\begin{equation}
\mathcal{Z}^{\star}_n= \int \mathcal{D}[\eta] e^{-N\mathcal{F}_n[\eta]}\, ,
\end{equation}
where $\mathcal{F}_n$ is the single vertex replicated free energy functional
\begin{equation}
\label{free}
\mathcal{F}_n[\eta]=\frac{K}{2e}\int d^n[\Xi] \eta^2(\Xi^{(n)})-\log \Bigg(\,\int d^n\Omega_{S} \bigg(\int d^n[\Xi]\eta(\Xi^{(n)}\,)\Phi(\Xi^{(n)},\S^{(n)})\,\bigg)^{K}\,\Bigg)\, 
\end{equation}
and the equilibrium free energy per particle $f^{\star}$, averaged on $\mathcal{G}^{\star}_{N,K}$, is given by
\begin{equation}
\label{lastsaddle}
\beta f^{\star}=\lim_{n\to 0}\,\,\partial_n \exp\Big(\,\min_{\eta}\,\mathcal{F}_n[\eta]\,\,\Big)\, .
\end{equation}

It can be proved that in the thermodynamic limit $f^{\star}$ converges to the free energy $f$, defined by \eqref{f}(assuming the thermodynamic limit exists). 

This result is a consequence of the fact that, in the large $N$ limit, the number of double edges and single self-loops is $\sim O(1)$ almost surely, whilst the probability that at least one more complicated local structure is present in whole the graph (for example double self-loops or triple edges) is $\sim O(1/N)$ \cite{Bollobas}. For this reason, we can switch from a given non-simple RRM, indicated by $G^{\star}$, to at least one simple RRG, indicated by $G$, changing only a finite set of different edges. We say that $G^{\star}$ and $G$ are almost equivalent to each other.

Since the Hamiltonians \eqref{ham} defined for two almost equivalent graphs $G^{\star}$ and $G$ differ to each other only for a finite number of couplings, the difference between the free energies per particle $f_{G^{\star}}$ and $f_{G}$ vanishes in the thermodynamic limit. For this reason, given a set $\mathcal{S}(G^{\star})\subset \mathcal{G}_{N,K}$ of simple $RRG$s, almost equivalent to the multigraph $G^{\star}$, obtained by a proper switching procedure, we have:
\begin{equation}
\label{graphproj}
f_{G^{\star}}\sim \frac{1}{|\mathcal{S}(G^{\star})|}\sum_{G\in \mathcal{S}(G^{\star})}f_G \,,
\end{equation}
so we obtain:
\begin{equation}
\label{graphproj2}
f^{\star}=\sum_{G^{\star}\in \mathcal{G}^{\star}_{N,K}}\mu^{\text{cm}} _{G^{\star}} f_{G^{\star}} \sim\sum_{G\in \mathcal{G}_{N,K}} \Bigg[\sum_{\left\{G^{\star};G\in \mathcal{S}(G^{\star})\right\} } \frac{\mu^{\text{cm}}_{G^{\star}}}{|\mathcal{S}(G^{\star})|}\,\,\Bigg ]f_G=\sum_{G\in \mathcal{G}_{N,K}} C^{\mathcal{S}}_{G}f_G\, ,
\end{equation}
where the sum $\sum_{\left\{G^{\star};G\in \mathcal{S}(G^{\star})\right\}} $ runs over all the multigraphs $G^{\star}$ such that $G\in \mathcal{S}(G^{\star})$ and $|\mathcal{S}(G^{\star})|$ is the cardinality of the set $\mathcal{S}(G^{\star})$. The quantity $\mu^{\text{cm}}_{G^{\star}}$ is the statistical weight of the multigraph $G^{\star}$ in configuration model, proportional to the reduced multplicity factor $M_{G^{\star}}$:
\begin{equation}
\mu^{\text{cm}}_{G^{\star}}=\frac{m_{G^{\star}}}{|\mathcal{G}_{N,K}|}=\frac{(K!)^N}{(KN-1)!!}m_{G^{\star}}
\end{equation}
If we use the switching procedure proposed by B.D. McKay and N.C. Wormald \cite{McKay} all factors $C^{\mathcal{S}}_{G}$ are equivalent, so we recover the quenched free energy defined on \eqref{f}.

A similar approach can be extended to all partition functions with the form described in \eqref{Zrep} and \eqref{P}.

An alternative way, still in replica formalism, to obtain a variational free energy for spin models on Bethe lattice has been derived by Mottishaw and De Dominicis for Ising spin-glass \cite{M}. The extension of the scheme proposed by \cite{M} to this model is quite easy. The cavity method also provides another possible method. The advantage of the variational approach based on the free energy functional \eqref{free} with respect to the other two cited methods is the fact that it offers an easier way to study the stability of the paramagnetic state. 

\section{Mean field equation and stability}
\label{3}
In order to find the global minimum of $\mathcal{F}_n[\eta]$, we must impose the stationary condition $\footnotesize {\delta \mathcal{F}_n[\eta]/\delta \eta[\Xi^{(n)}]=0}$, that provides the self-consistency mean field equation for the order parameter function $\eta$, in replica space. At this stage we formulate the problem for a generic integer value of $n$. At the end of this section we define the replica symmetric ansatz, that enables to compute the $n\to 0$ limit.

From the saddle point equation we obtain:
\begin{equation}
\label{RRG}
\eta(\Xi^{(n)})=e\frac{\int d^n\Omega_S\,\,\Big(\,\mathcal{U}[\eta]\big(\S^{(n)}\big)\,\Big)^{K-1}\Phi \big(\Xi^{(n)},\S^{(n)}\big) }{\int d^n\Omega_S\,\,\Big(\,\mathcal{U}[\eta]\big(\S^{(n)}\big)\,\Big)^K }\, .
\end{equation}
where the functional $\mathcal{U}[\eta]\big(\S^{(n)}\big)$ is given by
\begin{equation}
\mathcal{U}[\eta]\big(\S^{(n)}\big)\,=\int d^n[\Xi ]\eta\big(\Xi^{(n)}\,\big)\Phi(\Xi^{(n)},\S^{(n)})\, .
\end{equation}
It is noteworthy that, using just the basic relation \eqref{RRG}, we are able to compute directly the harmonic part:
\begin{multline}
\label{armonico}
\mathcal{F}_{\text{harm}}[\eta]=\frac{K}{2e}\int d^n[\Xi ] \eta (\Xi^{(n) } )^2\\=
\frac{K}{2}\frac{\int d^n[\Xi ]\eta (\Xi^{(n) } )\int d^n\Omega_S\,\,\Big(\,\mathcal{U}[\eta]\big(\S^{(n)}\big)\,\Big)^{K-1}\Phi(\Xi^{(n)},\S^{(n)}) }{\int d^n\Omega_S\,\,\Big(\,\mathcal{U}[\eta]\big(\S^{(n)}\big)\,\Big)^K }\\=
\frac{K}{2}\frac{\int d^n\Omega_S\,\,\Big(\,\mathcal{U}[\eta]\big(\S^{(n)}\big)\,\Big)^K }{\int d^n\Omega_S\,\,\Big(\,\mathcal{U}[\eta]\big(\S^{(n)}\big)\,\Big)^K }=\frac{K}{2}\, .
\end{multline}
This quantity does not depend on $n$ so it drops away when we compute the limit \eqref{lastsaddle}. 

The self-consistency equation suggests that a general solution has the form
\begin{equation}
\label{Xi:S}
\eta\big(\Xi^{(n)}\big)=\sqrt{e\,}\int d^n\Omega _S\,\, \rho(\S ^{(n)}) \Phi(\Xi^{(n)},\S ^{(n)})\,,
\end{equation}
where spin replicas distribution $\rho$ is solution of the equation
\begin{equation}
\label{1:rho}
\rho(\S ^{(n)})=\frac{\Big(\,\widetilde{\mathcal{U}}[\rho]\big(\S ^{(n)}\big)\,\Big)^{K-1}}{\int d^n\Omega_{S'}\,\,\Big(\,\widetilde{\mathcal{U}}[\rho]\big(\S '^{(n)}\big)\,\Big)^K }
\end{equation}
and
\begin{multline}
\label{U:rho}
\widetilde{\mathcal{U}}[\,\rho\,](\S ^{(n)})=\frac{1}{\sqrt{e\,}}\,\U\big[\eta[\rho]\,\,\big](\S ^{(n)})\\=
\bigg(\frac{\beta}{2\pi}\bigg)^{\frac{n}{2}}\int d^n[\Xi]\int d^n\Omega_{S'}\rho(\S '^{(n)}) \exp\bigg(\frac{\beta}{2}\sum_{a=1}^n\Big(Tr\big[(\Xi^a)^2\,\big]+ i \sum_{\beta,\gamma} \Xi_{\beta,\gamma}^a \big(S_{\beta}^aS_{\gamma}^a+S_{\beta}'^aS_{\gamma}'^a\big)\,\Big)\,\bigg)\\=
\int d^n\Omega_{S'}\rho(\S'^{(n)}) \exp\bigg(-\beta\sum_{a=1}^n \big(\S^a\cdot \S '^a\big)^2 \,\bigg)\, .
\end{multline}
From equation \eqref{1:rho} one can easily check that, for odd $K$, the distribution $\rho(\S^{(n)})$ is non-negative; from regularity argument, we can argue that this result should be true also for even $K$. 

The $\rho(\S^{(n)})$ is not a normalized distribution, so it is more convenient to reformulate the problem in term of the normalized distribution order parameter $\hat\rho(\S ^{(n)})$ (a probability distribution), that is solution of the new self-consistency equation
\begin{equation}
\label{rhobis}
\hat\rho(\S ^{(n)})=\frac{\Big(\,\widetilde{\mathcal{U}}[\hat{\rho}]\big(\S ^{(n)}\big)\,\Big)^{K-1}}{\int d^n\Omega_{S'}\,\,\Big(\,\widetilde{\mathcal{U}}[\hat{\rho}]\big(\S '^{(n)}\big)\,\Big)^{K-1} }\,
\end{equation}
and it is related to $\rho$ in this way:
\begin{equation}
\label{rho:rho}
\rho(\S ^{(n)})=\sqrt{\,\frac{\int d^n\Omega_{S'}\,\,\Big(\,\widetilde{\mathcal{U}}[\hat{\rho}]\big(\S '^{(n)}\big)\,\Big)^{K-1}}{\int d^n\Omega_{S'}\,\,\Big(\,\widetilde{\mathcal{U}}[\hat{\rho}]\big(\S '^{(n)}\big)\,\Big)^K }\,\,}\,\,\,\hat{\rho}(\S ^{(n)})\, .
\end{equation}
We have introduced two way to study this problem: the first one in term of the function $\eta$, depending on $n$ matrices $\Xi^{(n)}$, and the second one in term of the distribution $\rho$, depending on $n$ spins $\S^{(n)}$. In the next sections we will refer to these two representations as $\Xi$ representation and $\S$ representation: we can switch from one to the other using the relation \eqref{Xi:S} standing between $\eta$ and $\rho$.
\subsection{Stability}
Given a solution $\eta^{\star}(\Xi^{(n)})$ of the stationary equation \eqref{RRG}, we must verify if it corresponds to a minimum or just a saddle-point. If $\eta^{\star}(\Xi^{(n)})$ is a local minimum, small deviations of the order parameter from it are thermodynamically unfavored, so such solution is locally stable.

Proving that $\eta^{\star}(\Xi^{(n)})$ is the global minimum of the variational free energy is a very difficult task: the analysis presented in this paper is limited to the local level.

 Using the $\Xi$ representation, we put a “small", in some sense, arbitrary displacement $\delta \eta$ from the stationary solution: 
if the free energy functional value increases for every perturbation, $\mathcal{F}[\eta^{\star}]$ is a local minimum\cite{AT}. This requirement can be checked from the second order expansion of $\mathcal{F}[\eta^{\star}+\delta \eta]$ over the perturbation around $\eta^{\star}$:
\begin{equation}
\mathcal{F}_n[\eta^{\star}+\delta \eta]\sim\mathcal{F}_n[\eta^{\star}]+\frac{1}{2}\delta^2_{\delta \eta}\mathcal{F}_n[\eta]\big|_{\eta=\eta^{\star}}=\mathcal{F}_n[\eta^{\star}]+\frac{1}{2}\Braket{\delta \eta|\mathcal{K}_{\eta^{\star}}\delta\eta}\, ,
\end{equation}
where the second order therm is given by:
\begin{multline}
\label{2order}
\Braket{\delta \eta|\mathcal{K}_{\eta^{\star}}\delta\eta}=\int d^n[\Xi]d^n[\Xi'] \dfrac{\delta^2\mathcal{F}[\eta]}{\delta \eta(\Xi^{(n)})\delta \eta(\Xi'^{(n)})}\Bigg|_{\eta=\eta^{\star}}\delta \eta(\Xi^{(n)})\delta \eta(\Xi'^{(n)})\\=
\frac{K}{e}\int d^n[\Xi] \big(\delta\eta(\Xi^{(n)})\,\big)^2
+\Bigg[\,\,K\frac{\int d^n[\Xi]\int d^n\Omega_S\,\,\Big(\,\mathcal{U}[\eta^{\star}]\big(\S ^{(n)}\big)\,\Big)^{K-1}\,\Phi(\Xi^{(n)},\S ^{(n)})\delta \eta(\Xi^{(n)})}{\,\int d^n\Omega_S\,\,\Big(\,\mathcal{U}[\eta^{\star}]\big(\S ^{(n)}\big)\,\Big)^K \,}\,\,\Bigg]^2\\
-K(K-1)\frac{\int d^n[\Xi]d^n[\Xi']\int d^n\Omega_S\,\,\Big(\,\mathcal{U}[\eta^{\star}]\big(\S ^{(n)}\big)\,\Big)^{K-2}\delta \eta(\Xi^{(n)})\Phi(\Xi^{(n)},\S ^{(n)})\delta \eta(\Xi'^{(n)})\Phi(\Xi'^{(n)},\S ^{(n)})}{\,\int d^n\Omega_S\,\,\Big(\,\mathcal{U}[\eta^{\star}]\big(\S ^{(n)}\big)\,\Big)^K\, }\, .
\end{multline}
The symbol $\mathcal{K}_{\eta^{\star}}$ stands for the linear integral operator, the kernel of which is the second functional derivative of $\mathcal{F}$ evaluated on the saddle point function $\eta^{\star}$: the Hessian operator. 

In order to be sure that $\eta^{\star}$ is a minimum, the Hessian operator must be non-negative. The non-negativity condition of the Hessian must be verified also in the $n\to 0$ limit.

Using the self-consistency equation for $\eta^{\star}$, we can rewrite \eqref{2order} in a simpler form:
\begin{multline}
\label{2order*bis}
\Braket{\delta \eta|\mathcal{K}_{\eta^{\star}}\delta\eta}=
\frac{K}{e}\int d^n[\Xi] \big(\delta\eta(\Xi^{(n)})\,\big)^2
+\Bigg[\,\,\frac{K}{e}\int d^n[\Xi]\eta^{\star}(\Xi^{(n)})\delta \eta(\Xi^{(n)})\,\,\Bigg]^2\\
-K(K-1)\frac{\int d^n[\Xi]d^n[\Xi']\int d^n\Omega_S\,\,\Big(\,\mathcal{U}[\eta^{\star}]\big(\S ^{(n)}\big)\,\Big)^{K-2}\Phi(\Xi^{(n)},\S ^{(n)})\Phi(\Xi'^{(n)},\S ^{(n)})\delta \eta(\Xi^{(n)}) \delta \eta(\Xi'^{(n)})}{\,\int d^n\Omega_S\,\,\Big(\,\mathcal{U}[\eta^{\star}]\big(\S ^{(n)}\big)\,\Big)^K\, }\, .
\end{multline}

The Hessian can be written as a sum over three operators:
\begin{equation}
\mathcal{K}_{\eta^{\star}}=\frac{K}{e}I+K^2 P_{\eta^{\star}}-K(K-1)\mathcal{J}_{\eta^{\star}}\,.
\end{equation}
The first one is proportional to an identity operator and the second one is proportional to a projector on the stationary solution $\eta^{\star}$, while the third one is a more complicated operator.

It is simple to verify that each solution $\eta^{\star}$ is an eigenvector of its own stability operator $\mathcal{K}_{\eta^{\star}}$ with eigenvalue $2K/e$, the maximal one; in literature this kind of non-degenerate eigenvalue is usually referred as the longitudinal eigenvalue.

The solution $\eta^{\star}$ is always a stable eigenvector for its associated Hessian, so we can restrict our stability quest on the functions that are orthogonal to $\eta^{\star}$. The stability operator restricted to this subspace is:
\begin{equation}
\mathcal{K}^{\bot}_{\eta^{\star}}=\frac{K}{e}I-K(K-1)\mathcal{J}_{\eta^{\star}}\, .
\end{equation}
If we switch to the $\S$ representation the Hessian operator corresponds to
\begin{equation}
\small
\big[\mathcal{K}^{\bot}_{\rho^{\star}}\delta \rho\,\big]\,\,(\S^{(n)})=\frac{K}{e}\delta \rho(\S^{(n)})
-K(K-1)\frac{\,\,\Big(\,\mathcal{\widetilde{U}}[\rho^{\star}]\big(\S ^{(n)}\big)\,\Big)^{K-2}\int d^n\Omega_{S'}\,\,e^{-\beta\sum_{a=1}^n(\S^a\cdot\S'^a)^2}\delta \rho(\S'^{(n)})}{\,e\,\int d^n\Omega_{S'}\,\,\Big(\,\mathcal{\widetilde{U}}[\rho^{\star}]\big(\S '^{(n)}\big)\,\Big)^K\, }\, ,
\end{equation}
where the symbol $\big[\mathcal{K}^{\bot}_{\rho^{\star}}\delta \rho\,\big]$ denotes the linear operator $\mathcal{K}^{\bot}_{\rho^{\star}}$ acting on the perturbation function $\delta \rho$ and returning a function depending on $\S^{(n)}$. 
\subsection{Replica symmetric mean field equation}
As usual in the replica formalism, the replica limit $n\to 0$ implies some ambiguities for the free energy functional \eqref{free} and the self consistency equation \eqref{rhobis}, indeed a function depending on a non-integer number of variables is somewhat meaningless. In order to compute the replica limit, some ansatz must be imposed. In this section we present the replica symmetric ansatz.

At the local level, typical random regular graphs look locally homogeneous tree like structures (no fluctuations of connectivity and interactions), so all such graphs seem equivalent on finite length scale. The random character of regular graphs comes from the large-scale loops contribution. If a single state exists, we can invoke the clustering property, i.e. the correlation function between two non directly interacting spins vanishes. Thanks to this property we can argue that contributions from large-scale loops vanishes in the $N\to\infty$ limit for almost every regular graph: the partition functions for all these graphs are the same, so the annealed and quenched averages over the graphs coincide. 
For this reason, single vertex functions, such as $\rho(\S^{(n)})$, are factorized in $n$ single replica functions:
\begin{equation}
\label{repsim}
\hat{\rho}(\S^{(n)})=\prod_{a=1}^n r(\S^a)\, .
\end{equation}
The distribution $r(\S)$ is the replica symmetric order parameter.

Imposing the hypothesis \eqref{repsim} in the equation \eqref{1:rho}, we obtain the replica symmetric mean-field equation
\begin{equation}
\label{RSeq}
r(\S)=\frac{\big(\,\widetilde{u}[r]\big(\S \big)\,\big)^{K-1}}{\int d^n\Omega_{S'}\,\,\big(\,\widetilde{u}[r]\big(\S '\big)\,\big)^{K-1} }\, ,
\end{equation}
where
\begin{equation}
\widetilde{u}[r](\S)=\int d\Omega_{S'}\, r(\S')e^{-\beta(\S\cdot \S')^2}\, .
\end{equation}
For a given solution $r^{\star}(\S)$, the equilibrium free energy per particle is given by:
\begin{equation}
f(\beta)=\frac{1}{\beta}\Bigg[\frac{K-2}{2}\log \bigg(\,\int d^n\Omega_{S'}\,\,\big(\,\widetilde{u}[r^{\star}]\big(\S '\big)\,\big)^K\,\bigg)-\frac{K}{2}\log \bigg(\,\int d^n\Omega_{S'}\,\,\big(\,\widetilde{u}[r^{\star}]\big(\S '\big)\,\big)^{K-1}\,\bigg)\,\,\Bigg]\,.
\end{equation}

The replica symmetric order parameter describes the probability density of the orientation of a single spin, averaged over the ensemble of random regular graphs:
\begin{equation}
r(\S)=\overline{\Braket{\,\mathfrak{r}(\S,\{\S_i\})\,}_{G}}\, ,
\end{equation}
where $\mathfrak{r}(\S,\{\S_i\})$ is the microscopic spin density:
\begin{equation}
\mathfrak{r}(\S,\{\S_i\})=\frac{1}{N}\sum_{i=1}^N \delta^D(\S-\S_i)
\end{equation}
and $\Braket{\cdot}_{G}$ is the thermal average for a given graph.

Note that, since the order parameter is a single spin distribution, the system must be homogeneous, i.e. the orientations of the spins in different sites are identical distributed random variables, with distribution $r(\S)$ (almost surely): 
\begin{equation}
\label{homogeneity}
\frac{1}{N}\sum_{i=1}^N \left(\,\overline{\Braket{\,\delta^D(\S-\S_i)}_G}\,\right)^2=\left(\,\,\frac{1}{N}\overline{\sum_{i=1}^N\Braket{\,\delta^D(\S-\S_i)\,}_G}\,\,\right)^2\, .
\end{equation}
This is a direct consequence of the assumption \eqref{repsim}, indeed the presence of non trivial local fluctuations on the spin distributions cannot be encoded in a single spin distribution order parameter, such as $r(\S)$. In order to avoid such limitation, more generic ansatz than the replica symmetric one should be considered: a replica symmetry breaking formulation must be explored.
\section{Ergodic solution, free energy and stability}
\label{4}
In this section we compute the free energy per particle in the ergodic phase, i.e. the phase where a single pure state, the Gibbs one, occurs. From the linear stability analysis, we provide a necessary, but not sufficient, validity condition of the ergodicity assumption. We detect the presence of a critical temperature, depending on the connectivity, below which the ergodic phase solution is unstable. We can also argue about the nature of such instability.
\subsection{Ergodic solution}
In the ergodic phase, replica and rotation symmetries are not broken, so we can use the equation \eqref{RSeq}, and $r(\S)$ is a constant. Imposing this ansatz we obtain
\begin{equation}
r(\S)=\frac{\Gamma\big(\frac{D}{2}\big)}{2\pi^{\frac{D}{2}}}=\frac{1}{\Omega (D)}\, .
\end{equation}
The symbol $\Omega (D)$ stands for the surface of the unit hypersphere on $D$-dimensional space.
By analogy with magnetic systems, we shall call this $O(D)$-symmetric solution \textit{paramagnetic}.

The free energy per particle is given by:
\begin{equation}
f(\beta)=\frac{K-2}{2\beta}\log\big(\,\,\Omega (D)\,\,\big)-\frac{K}{2\beta}\log\bigg(\,\Omega (D-1)\int d\theta \sin^{D-2}(\theta) \exp\big(\,-\beta \cos^2(\theta)\,\,\big)\,\bigg)\, .
\end{equation}

The kernel of the restricted Hessian operator, in $\Xi$ representation, is given by:
\begin{equation}
\label{var2para}
K^{\bot}_{\eta_{\text{para}}}(\Xi^{(n)},\Xi'^{(n)})=\frac{K}{e}\delta(\Xi'^{(n)}-\Xi^{(n)})-\frac{K(K-1)}{e\,A_{0}^{n}(\beta)} \int d^n\Omega_S\, \Phi \big( \Xi^{(n)},\S ^{(n)}\big) \Phi \big( \Xi'^{(n)},\S ^{(n)} \big)\,, 
\end{equation}
where
\begin{equation}
\label{C2}
A_{0}(\beta)=\Omega (D-1)\int d\theta _1 \sin^{D-2}(\theta_1)\,\, \exp \bigg( -\beta \cos^2(\theta_1)\,\,\bigg)\, .
\end{equation}
It can be shown that the eigenvectors are all functions with the form:
\begin{equation}
\label{autovet}
\eta_{2l_1\{\mu\}_1;2l_2\{\mu\}_2;\cdots ;2l_n,\{\mu\}_n}(\Xi^{(n)})=\sqrt{e\,}\int d^n\Omega_S\, \, \bigg(\prod_{a=1}^n i^{l_a} Y^{(D)}_{2l_a\{\mu\}_a}(\S^a)\bigg)\Phi ( \Xi^{(n)},\S ^{(n)})\, ,
\end{equation}
where $Y^{(D)}_{l,\{\mu\}}(\S)$ is the hyperspherical harmonic with indices $\left(\,l\,\{\mu\}\,\right)$, calculated in angular coordinates associated to $D$-component spin $\S$. The imaginary factor $i^{l_a}$ is due to the fact that the functions $\eta_{2l_1\{\mu\}_1;\cdots ;2l_n,\{\mu\}_n}$ must be real.
 
For each integer $l$, we define the quantity 
\begin{equation}
\label{autovalori}
 A_{2l}(\beta)=\frac{(4l+D-2)}{(D-2)\,\omega^D_{2l}}\Omega(D-1)\int d\theta \sin^{D-2}(\theta) e^{-\beta \cos^2(\theta)}C^{\frac{D-2}{2}}_{2l}\big(\cos (\theta)\,\big)\, ,
\end{equation}
where
\begin{equation}
\omega^D_{2l}=\frac{(D+4l-2)(2l+D-3)!}{(2l)!(D-2)!}
\end{equation}
and the function $C^{\frac{D-2}{2}}_{2l}(x)$ is the Gegenbauer polynomial \cite{Avery}:
\begin{equation}
\label{gegenbauer}
C^{\frac{D-2}{2}}_{2l}(x)=\frac{1}{(D-4)!!}\sum_{t=0}^l\dfrac{(-1)^t(D+4l-2t-4)!!}{(2t)!!(2l-2t)!}\,x^{2l-2t}\, .
\end{equation}

The two lowest eigenvalues are:
\begin{gather}
\label{lambda22}
\Lambda_{22}(\beta,K)=\frac{K}{e}-\frac{K(K-1)}{e}\bigg(\frac{A_{2}(\beta)}{A_0(\beta)}\bigg)^2\, ,\\
\label{lambda4}
\Lambda_{4}(\beta,K)=\frac{K}{e}-\frac{K(K-1)}{e}\frac{A_{4}(\beta)}{A_0(\beta)}\, .
\end{gather}
The associate eigenfunctions, in $\S$ representation, are respectively
\begin{gather}
\label{rho22}
\rho_{22}(\S^{(n)})\propto Y_{2,\{\mu\}}(\S^a)Y_{2,\{\mu '\}}(\S^b)\, ,\\
\label{rho4}
\rho_{4}(\S^{(n)})\propto Y_{4,\{\mu\}}(\S^a)\, ,
\end{gather}
for some choice for $\{\mu\}$, $\{\mu '\}$ and replica indices.

From the eigenvalues \eqref{rho22} and \eqref{rho4} we can obtain two \textit{critical} temperatures $T_{22}(K)$ and $T_{4}(K)$, for any given connectivity $K$, satisfying the conditions:
\begin{gather}
\label{Tcrit1}
\frac{1}{K-1}=\bigg(\frac{A_{2}(\beta_{22})}{A_0(\beta_{22}\,)}\bigg)^2\, ,\\
\label{Tcrit2}
\frac{1}{K-1}=\frac{A_4(\beta_4\,)}{A_0(\beta_4)}\, .
\end{gather}
The highest temperature controls the instability and the other one has no direct physical meaning; the true critical temperature curve is given by:
\begin{equation}
\label{Tcrit}
T_c(K)=\max \big(\,T_{22}(K),\,T_{4}(K)\,\big)\, .
\end{equation}

The character of the instabilities described by the two lines can be guessed by general argument.

 For $T<T_c(K)$ a new, locally stable, solution $\rho'(\S^{(n)})$ bifurcates continuously from the unstable replicated paramagnetic solution $\rho_{\text{p}}(\S^{(n)})$. Near the critical temperature, we can consider the difference between these two solutions as a perturbation:
\begin{equation}
\rho'(\S^{(n)})=\rho_{\text{p}}(\S^{(n)})+\delta \rho(\S^{(n)})\, .
\end{equation}
Let us suppose that the distribution $\rho'$ is replica symmetric. We can pass to the normalized distribution $\hat{\rho}'$ through the relation \eqref{rho:rho} and use the ansatz \eqref{repsim}:
\begin{equation}
\hat{\rho}'(\S^{(n)})=\prod_{a=1}^n r'(\S^a)=\prod_{a=1}^n\big( r(\S^a)+\delta r(\S^a)\,\big)\sim \hat{\rho}_{\text{p}}(\S^{(n)})+\sum_{a=1}^n \delta r(\S^a)\, ,
\end{equation}
that implies
\begin{equation}
\label{PAFdeviation}
\delta \rho(\S)\propto \sum_{a=1}^n \delta r(\S^a)\, .
\end{equation}
For $T$ close enough to $T_c$, the deviation from ergodic solution is basically dominated by the critical eigenvector. 

If $T_c(K)=T_{4}(K)$, the critical eigenvector \eqref{rho4} has the form \eqref{PAFdeviation}, depending only on a single replica's variables, then the system is locally stable with respect to the replica symmetry breaking and a simple $O(D)$ symmetry breaking occurs. In this case, the paramagnetic solution is unstable toward the appearance of an anti-ferromagnetic order. 

If $T_c(K)=T_{22}(K)$, the critical eigenvector \eqref{rho22} depends on two different replicas, so the system breaks the replica symmetry.

By analogy with the magnetic models, we shall refer to the first case as paramagnetic/anti-ferromagnetic instability (P-AF) and to the second case as paramagnetic/spin-glass instability (P-SG).

The stability of the paramagnetic solution exclude the presence of a continuous phase transition at $T>T_c(K)$, for any fixed connectivity $K$. However the system may undergo a discontinuous transition, so the critical temperature $T_c(K)$ may not correspond to a real transition temperature, but provides a lower bound to it.

\subsection{Instability} 
The conditions \eqref{Tcrit1},\eqref{Tcrit2} and \eqref{Tcrit} define an instability line on the temperature-connectivity plane, that separates the high-T/low-K region, where the paramagnetic solution is locally stable, from the low-T/high-K region, where the paramagnetic solution is unstable. 

\begin{figure}[h!]
\center
{\includegraphics[scale=0.6]{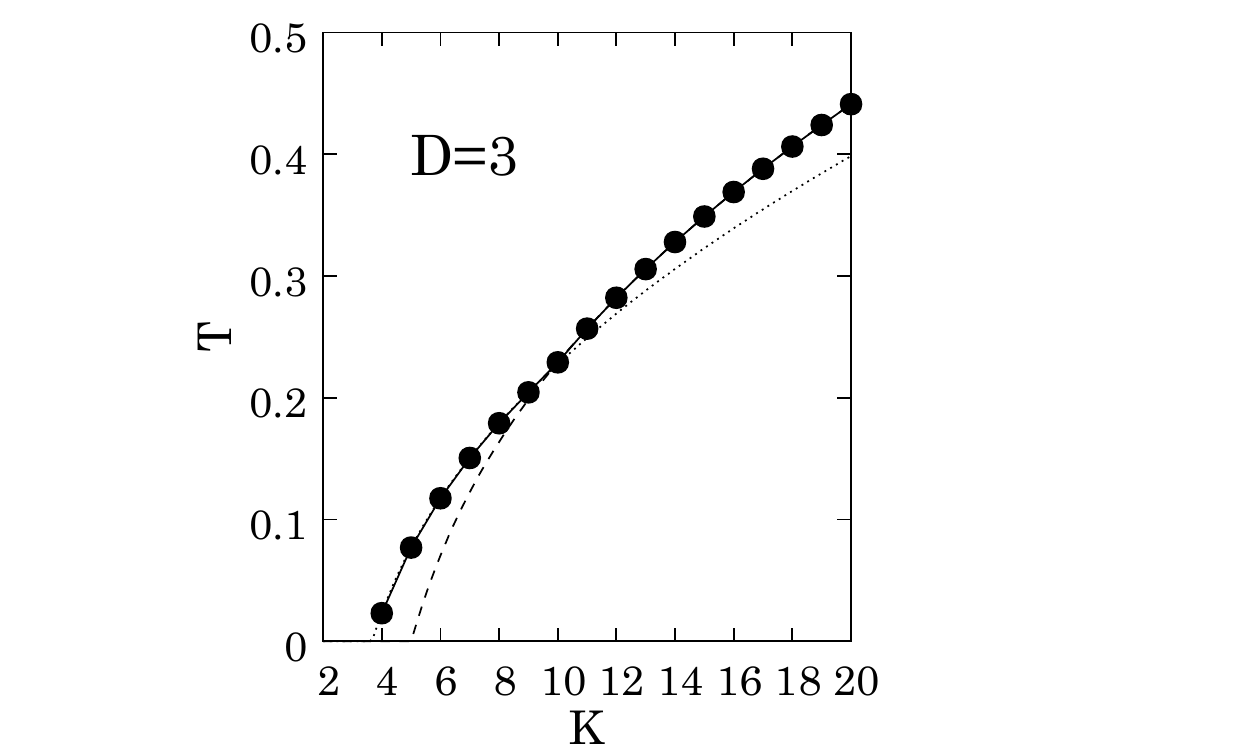}}
\hskip -3.9cm
{\includegraphics[scale=0.6]{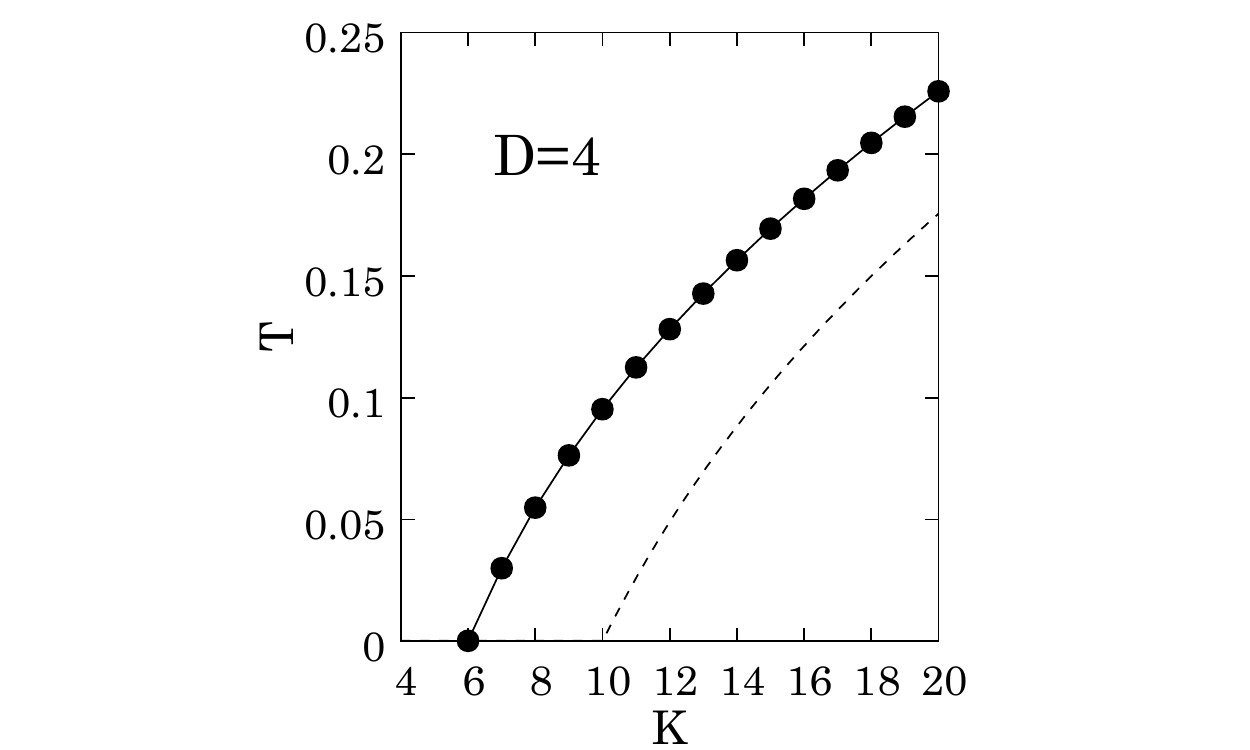}}
\hskip -3.4cm
{\includegraphics[scale=0.6]{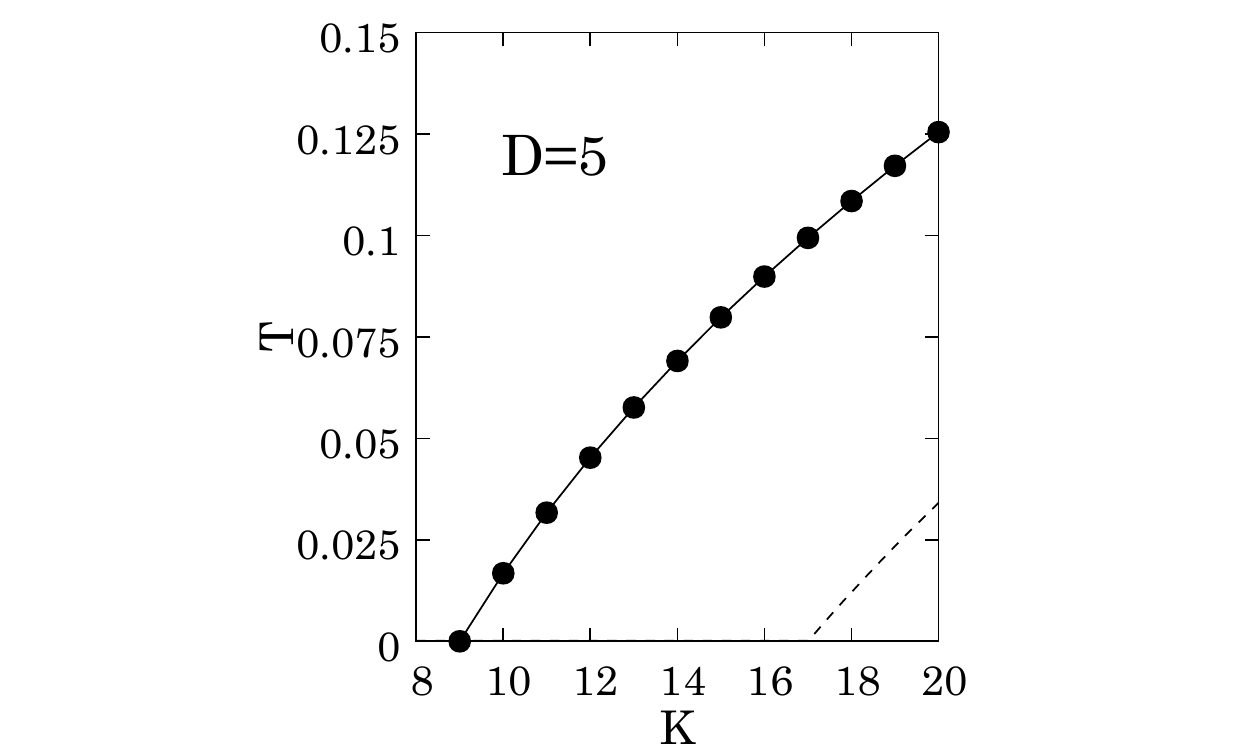}}
\caption{\small Instability lines for $D=3$, $D=4$ and $D=5$ in temperature-connectivity plane. The black points are the critical temperature values $T_c(K)$ and black solid line is the extension to real $K$, whilst dashed line is the continuation of $T_{22}(K)$ and dotted line at $D=3$ is the continuation of $T_{4}(K)$. The temperature $T$ is in unit of the coupling constant over the Boltzmann constant.}
\label{fig:subfig}
\end{figure}

\begin{figure}[h!]
\centering
{\includegraphics[width=0.7\textwidth]{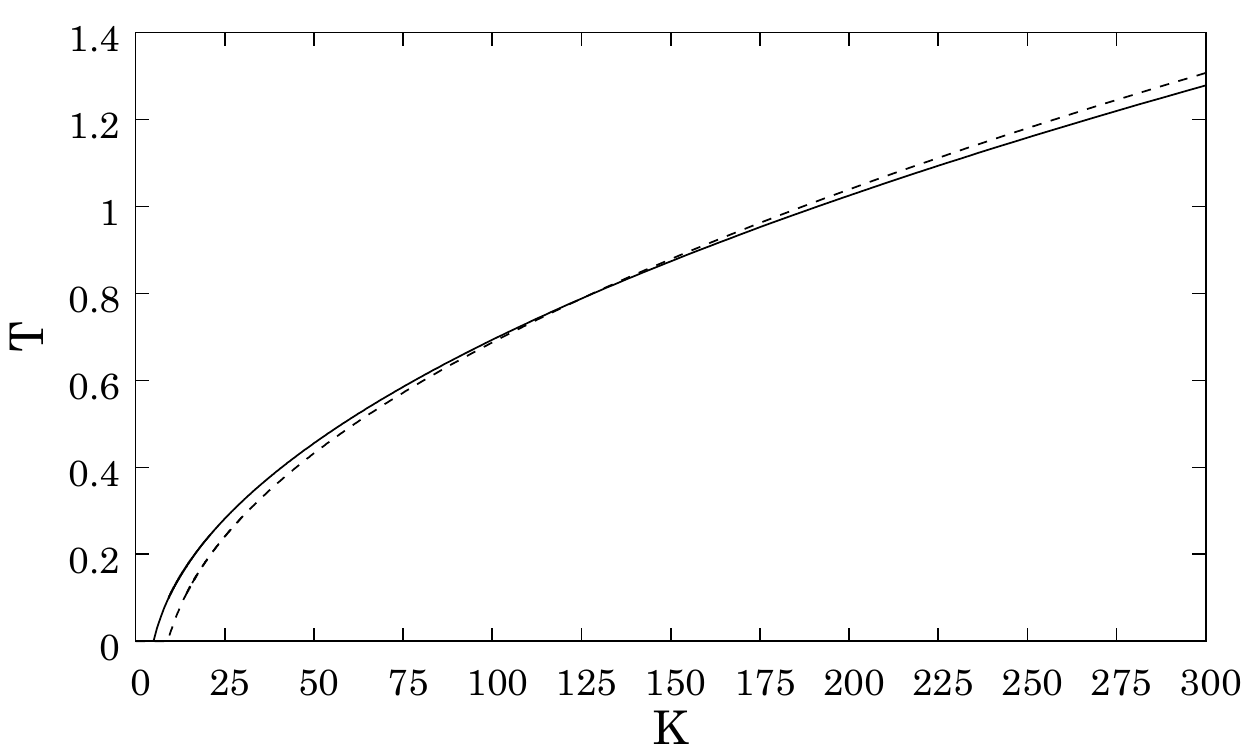}} \quad
\caption{\small Instability lines for $D=4$ in temperature-connectivity plane up to $K=300$. Solid and dashed lines are the extension to real $K$ respectively for $T_{4}(K)$ and $T_{22}(K)$.}
\label{fig:subfig2}
\end{figure}

The instability lines for $3$,$4$ and $5$ components spins are plotted, up to $K=20$, in figure \ref{fig:subfig} and for $D=4$, up to $K=300$ in figure \ref{fig:subfig2}. Critical temperatures for several connectivities are reported in tables \ref{temp}.


 \begin{table}[h!]
\small
\centering
\begin{tabular}{lccc}
D=3\\
\\
\toprule
$K$ & $T_c \,(J/k_b)$\\
\midrule
3& $-$ \\
4 & $0.023161$ \\
5 & $0.077058$ \\
6 &$0.117442$\\
7 &$0.150509$\\
8&$0.179054$l\\
9&$0.204506$\\
10&$0.229143$\\

12&$0.282114$\\

14&$0.327916$\\

16&$0.368909$\\

18&$0.406378$\\

20&$0.441121$\\
\bottomrule
\\
\end{tabular}
\qquad\quad
\begin{tabular}{lccc}
D=4\\
\\
\toprule
$K$ & $T_c \,(J/k_b)$\\
\midrule
3& $-$ \\
4 & $-$ \\
5 & $-$ \\
6 & $-$\\
7 &0.029984\\
8&0.054885\\
9&0.076334\\
10&0.076333\\

12&0.128199\\

14&0.156501\\

16&0.181700\\

18&0.204632\\

20&0.225820\\
\bottomrule
\\
\end{tabular}\qquad\quad
\begin{tabular}{lccc}
D=5\\
\\
\toprule
$K$ & $T_c \,(J/k_b)$\\
\midrule
3& $-$ \\
4 & $-$ \\
5 & $-$ \\
6 & $-$\\
7& $-$\\
8&$-$\\
9&$-$\\
10&0.016767\\

12&0.045232\\

14&0.069094\\

16&0.089882\\

18&0.108482\\

20&0.125444\\
\bottomrule
\\
\end{tabular}\\
\caption{\small Critical temperatures at several $K$s for $D=3$, $D=4$ e $D=5$. }
\label{tabella}
\label{temp}
\end{table}

As in the anti-ferromagnetic Potts model, there exists a lower critical connectivity $K_L>2$, below which the paramagnetic solution is stable also at $T=0$, despite the existence of an extensive number of loops.

At $K=K_L$, the paramagnetic solution is unstable toward the appearance of an anti-ferromagnetic order (P-AF instability).
The value of $K_L$ can be computed analytically:
\begin{equation}
\label{K_L}
\frac{1}{K_L-1}=\lim_{\beta \to 0}\, \frac{A_4(\beta)}{A_0(\beta)}\longrightarrow K_L=\frac{D^2+2}{3}\, .
\end{equation}
Note that, for $D\leq11$, the connectivity $K_L$ is lower than the connectivity where AF $D$-PM, defined on the same ensemble of graph \cite{Kraz}, undergoes a discontinuous dynamical transition. It is reasonable to assume that, in this case, no discontinuous transitions occur at $K<K_L$ and there may exist a range of connectivities $K\geq K_L$ where $T_c(K)$ properly a transition temperature.

By analogy with AF-$3$-PM, we also guess that, for $D=3$, no discontinuous transitions occur at all and the instability line $T_c(K)$ provides the right phase diagram in the whole temperature-connectivity plane. A rigorous proof of this hypothesis is needed.

We stress that the P-AF instability does not correspond to the modulation instability \cite{Bir}, observed in AF $D$-PM by Zdeborov\'a and Krzakala with the Bethe-Peierls (BP) approach \cite{Kraz}.

As is well known, in the BP approach, the RS free energy is given by the fixed point of recursive equations, defined through the graph \cite{MM}. Actually, the modulation instability is an instability of the paramagnetic fixed point under BP iterations \cite{Bir}. 

Recasting the present model in the BP formalism, it can be shown that the stability criterion of the paramagnetic fixed point reads: 
\begin{equation}
\label{modulation}
(K-1)\left|\frac{A_{22}(\beta)}{A_{0}(\beta)}\right|\geq1\,.
\end{equation}
The lower connectivity $K_m$ where such instability appears (modulation connectivity) is:
\begin{equation}
\label{K_L}
\frac{1}{K_m-1}=\lim_{\beta \to 0}\, \left|\frac{A_{22}(\beta)}{A_0(\beta)}\right| \longrightarrow K_m=D,
\end{equation}
that coincides with the value obtained in AF $D$-PM.

If we consider the present model defined on a Cayley tree, at $K=K_m$ the paramagnetic fixed point turns to be unstable toward the appearance of an anti-ferromagnetic order, where, at each site, the corresponding spin has a single privileged orientation. In this case, it is possible to choose boundary conditions such that the distribution of the spins'orientations, through the graph, is periodic (crystal phase \cite{Bir}). However, in $RRG$, as discussed in \cite{Kraz} and \cite{Bir}, such kind of solutions are incompatible with the presence of frustrated loops, so the modulation instability is prevented and the Gibbs state is still extremal. 

The appearance of such instability is an artifact of the $BP$ approach. The BP approach turns to be exact only for Cayley tree, and its validity can be extended to generic sparse graphs only under some conditions. 

By contrast, the method proposed in this paper works directly with multigraph, and the presence of frustrating loops is explicitly encoded in the variational free energy. As a consequence, the modulation instability is automatically suppressed, indeed the Hessian \eqref{var2para}
 is positive-definite at $K=K_m$.

Obviously, the extremal solution of the RS saddle-point equation \eqref{RSeq} and the BP fixed-point are equivalent. The difference arises when we consider deviations around such saddle-point/fixed-point.

As discussed in the previous section, the P-AF instability, at $K=K_L$, should announce a transition to a replica symmetric phase, where the $O(D)$ symmetry is broken. 

The solution previously described, arising in Cayley tree at $K=K_m$, is avoided in RRG, so this phase must describe another kind of anti-ferromagnetic order. 

\begin{figure}[H]
\centering
{\includegraphics[width=0.9\textwidth]{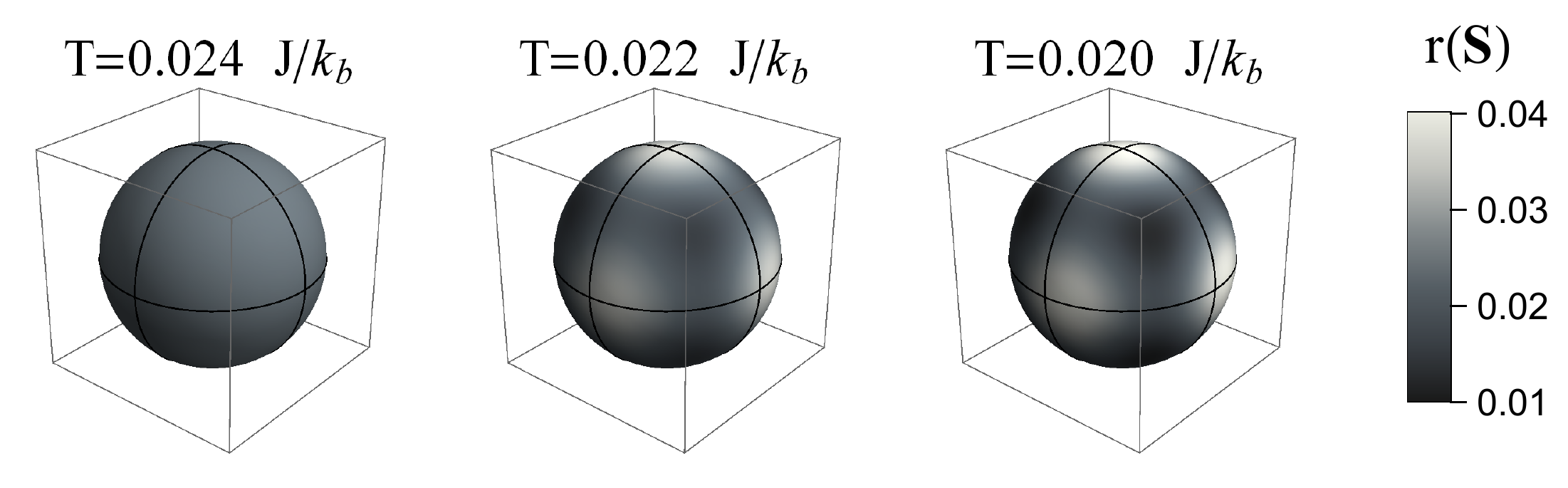}} 
\caption{\small Approximated single spin distribution for $D=3$ and $K=4$, at temperature higher than $T_c(4)$(left) and lower (the others). The value of $r(\S)$ assumed on the unit sphere is represented in grayscale colors.}
\label{figsfe}
\end{figure}

We guess that, in this phase, each state is characterized by $D$ privileged orthogonal axes. Because the presence of loops, the averaged single spin probability distribution $r(\S)$ must be symmetric under permutation of such axes, but is not uniform through the whole unit sphere, since the $O(D)$ symmetry breaking, and has equivalent maxima along the privileged axes (fig. \ref{figsfe}). 
 
In the paramagnetic phase, each spin is correlated only to its nearest neighbors, so it can rearrange paying a little energy coast. For this reason, each spin can spread uniformly through the unit sphere. In this anti-ferromagnetic phase, long-range correlations are present, since the high number of constraint due to the first neighbor interactions. For this reason, each spin drifts along some most likely directions. Actually, this instability does not have a discrete counterpart in AF $D$-PM.
 
An approximated evaluation of the anti-ferromagnetic solution, near $T_c(K)$, is performed at $K=4$ and $D=3$. We consider distributions $r(\S)$ of the form:
\begin{equation}
r(\S)=\frac{1}{4\pi}+\sum_{m=-4}^4a_mY_{4,m}(\S)\,,
\end{equation}
where $Y_{4,m}(\S)$ are the three dimensional spherical harmonics corresponding to the critical eigenvectors \eqref{rho4}. The RS free energy is minimized with respect to the parameters $a_m$. The solution is reported in graphics \ref{figsfe}, for several temperature, and it is in agreement with our qualitative expectation.

As in every $O(D)$, symmetry breaking theories, all the anti-ferromagnetic solutions of this kind are equivalent under a proper rotation of the reference frame.

Note that the lower connectivity with a non zero $T_{22}(K)$ is $K=(D-1)^2+1$, that remarkably coincides with the rigorous upper bound limit to the paramagnetic extremality condition of \cite{Stigum} for AF $D$-PM, rederived in\cite{Kraz} from the divergence of the spin glass susceptibility. 

For $D=3$, the $T_{4}(K)$ line and $T_{22}(K)$ line cross at $K=10$, so the system has a P-SG instability for larger connectivities; the two lines also cross for $D=4$, around $K=124$ (fig. \ref{fig:subfig2}). 

Because of the high value of the connectivity, we suppose that, at $D=4$, such cross is more likely to be prevented by the appearance of a first-order transition at a lower connectivity. 

By contrast, for $D=3$, we guess that the instability line corresponds to a proper continuous P-SG transition at $K>10$. Since the replica symmetry breaking, the system can no longer verify the condition \eqref{homogeneity}, i.e. it cannot relax into a phase where the distributions of the spin's orientation in each site are the same, as in the anti-ferromagnetic and the paramagnetic phases. By analogy with others spin glass models, this phase should be dominated by the presence of a large number of equivalence classes of states\footnote{We say that two states are equivalent if one can be transformed into the other by a rotation of the reference frame in which the spins are represented}, each of which described by a set of $N$ different single spin random correlated distributions, corresponding to each site of the graph.

Since the system shows a discontinuous changing in the character of the instability at high connectivities, the thermodynamic behavior at low connectivities (near $K_L$) cannot be grabbed by an expansion around the $K\to\infty$ limit, as performed in \cite{G2} for Ising spin glass.

The asymptotic behavior of $T_{4}(K)$ and $T_{22}(K)$ for large $K$ ($K\gg K_L$) is derived, for a generic value of $D$, performing the power expansion in $\beta=1/T$ of the eigenvalues \eqref{lambda22} and \eqref{lambda4}, up to the fourth order, and than reverting the two expansion according respectively to the equations \eqref{Tcrit2} and \eqref{Tcrit1}.We obtain:
\begin{equation}
T_{22}(K)=\frac{2\sqrt{K-1}}{D (D+2) }+\frac{4-2 D}{D^2+4 D}-\frac{6 D(D+1) }{ (D+4)^2 (D+6)}\frac{1}{\sqrt{K-1}}+O\left(\frac{1}{K-1}\right)\,;
\end{equation}
\begin{multline}
T_4(K)=\frac{2 \sqrt{3(K-1)}}{\sqrt{D (D+2) (D+4) (D+6)\,} }+
\frac{4-2 D}{D^2+8 D}\\-\frac{2 \sqrt{ (D+4) (D+6)} ( 4 D^3+ 9 D^2 + 36 D -40 ) }{\sqrt{3D^3(D+2)\,} (D+8)^2 (D+10)}\frac{1}{\sqrt{K-1}}+O\left(\frac{1}{K-1}\right)\, .
\end{multline}
From this expansion we observe that the two \textit{critical} lines never cross for $D\geq 5$, so we have a P-AF instability for all value of $K$, so the instability pattern changes considerably from $D=3,4$ to higher number of components. No big qualitative differences appear varying the number of spin's components for $D\geq5$.

We remark that, in order to achieve a complete description of the replica symmetric phase, the anti-ferromagnetic solution must be computed. This problem may be tackled via the low-temperature expansion of the equation \eqref{RSeq} and will be investigated in next works.

\section{Conclusion}
In this paper we study a model of $D$-components vector spins, lying on the vertices of a RRG, with first neighbor interactions $(\S_i\cdot\S_k)^2$. This model could be considered as a continuum version of Potts anti-ferromagnetic model.

The average over RRG ensemble is performed, via the replica trick, using a new variational approach, presented in this paper for the first time. Such method offers a simpler way to study the linear stability of replica symmetric solution, than other methods commonly used on sparse graphs.

The paramagnetic free energy is computed and the linear stability analysis is performed, studying the positivity of the second variation of the variational free energy around the paramagnetic saddle point. We obtain the instability line on the connectivity-temperature plane.

At low connectivity, the paramagnetic solution is stable also at zero temperature, up to a lower critical connectivity $K_L=(D^2+2)/3$, where the system undergoes a continuous instability toward another replica symmetric solution that breaks the $O(D)$ symmetry (anti-ferromagnetic solution). 

The instability line assures the presence of a phase transition. However, as we stressed, we cannot exclude the presence of a first-order phase transition: a non local analysis of the stability of the paramagnetic solution for this model is needed. This is a formidable task, also via a numerical approach.

Comparing this model with the anti-ferromagnetic $D$-states Potts model, we argued that the instability line should describe the proper phase diagram for $D=3$ at all connectivities, and also for $4\leq D\leq11$ near $K_L$. The possibility that the system undergoes a discontinuous transition should be explored.

We remark that, in order to achieve a complete description of the replica symmetric phase, the anti-ferromagnetic solution must be computed. 

It is also noted that the instability lines, for spins with $3$ and $4$ components, is completely different from what obtained for higher numbers of components. In these two cases, the paramagnetic phase manifests an instability with respect to replica symmetry breaking at high connectivities. 

The results, described in this paper, lay the necessary groundwork for next studies about finite connectivity Heisenberg glass models. 

We stress that the variational method, proposed in this paper, could be easily extended to many other systems with a finite connectivity and may represent a valid alternative to the cavity method. The RSB formulation within this approach will be investigated in next works.
\section*{Acknowledgement}
We are grateful to Giorgio Parisi and Federico Ricci Tersenghi for interesting discussions.

\end{document}